\documentstyle[twocolumn,aps,prl]{revtex}
\newcommand{\fcyc}{\vec{\omega}_c}

\begin{document}
\draft
\title{The two electron artificial molecule}
\author{B. Partoens,$^1$\cite{bart} A. Matulis,$^2$\cite{algis} and F. M.
Peeters$^1$%
\cite{francois}}
\address{$^1$Departement Natuurkunde, Universiteit Antwerpen (UIA),
Universiteitsplein 1, B - 2610 Antwerpen, Belgium \\
$^2$ Semiconductor Physics Institute, 
Go\v{s}tauto 11, 2600 Vilnius, Lithuania}
\date{\today}
\maketitle

\begin{abstract}
Exact results for the classical and quantum system of two vertically coupled
two-dimensional single electron quantum dots are obtained as a function of
the interatomic distance $(d)$ and with perpendicular magnetic field. The
classical system exhibits a second order structural transition as a function
of $d$ which is smeared out and shifted to lower $d$ values in the quantum
case. The spin-singlet $\leftrightarrow $ spin-triplet oscillations are
shifted to larger magnetic fields with increasing $d$ and are quenched for
a sufficiently large interatomic distance.
\end{abstract}


%%%%%%%%%%%%%%%%%%%%%%%%%%%%%%%%%%%%%%%%%%%%%%%%%
%			ABSTRACT		%
%%%%%%%%%%%%%%%%%%%%%%%%%%%%%%%%%%%%%%%%%%%%%%%%%

%\section{Introduction}

Quantum dots are {\it artificial atoms} which have been a subject of intense
theoretical and experimental research in recent years\cite{j95}. 
A very recent development is the study of vertically
coupled quantum dots\cite{pal95} or {\it artificial molecules}. In Ref.~%
\onlinecite{ima96} a double dot system containing three spin-polarized
electrons was investigated. A sequence of angular momentum magic numbers was
found which depends on the strength of the interdot tunneling. An
analytically solvable model of an arbitrary number of vertically coupled
quantum dots containing each two electrons but with a $1/r^2$
electron-electron interaction and with neglect of the interdot tunneling was
presented in Ref.~\onlinecite{bj95}. The first experimental realization of
vertically coupled quantum dot structures were recently 
reported\cite{aus98,stmh96}.

Two vertically coupled {\it classical} dots were investigated by two of
the present authors as a
function of the interatomic distance ($d$)\cite{psp97}. We found first and
second order transitions which are similar to structural phase transitions.
%The ground-state and eigenmodes of classical artificial molecules consisting
%of identical atoms containing up to ten electrons each were investigated.
For first (second) order transitions the first (second) derivative of the
energy with respect to $d$ is discontinuous, the radial position of the
particles changes discontinuously (continuously), and the frequency of the
eigenmodes exhibits a jump (one mode becomes soft, i.e., its frequency
becomes zero). Here we will investigate the influence of quantum effects on
the above structural transitions. As an example we limit ourselves to the
exact solvable artificial molecule with in each dot one electron. In the
classical limit this
system exhibits a second order transition. We also
investigate how the spin-singlet $\leftrightarrow $ spin-triplet
oscillations are modified by changing the interatomic distance. The
equivalent two electron quantum dot, i.e. $d=0$, was solved in
Refs.~\onlinecite{mhw91,wmc92,dhc92}. We will show that the inter-layer distance
introduces an additional parameter with which the electron-electron
interaction can be altered, and which modifies the behavior of the system
considerably.

%%%%%%%%%%%%%%%%%%%%%%%%%
%	MODEL		%
%%%%%%%%%%%%%%%%%%%%%%%%%
%\section{The model}
The present system consists of two layers, separated in the $z$-direction by a
distance $d$, with in each layer ($xy$-plane) one electron which is confined by
a parabolic potential. We take the same confinement strength in both layers and 
the barrier between the two layers is taken sufficiently large such that there 
is no tunneling between them.
Within the effective 
mass approximation, this system is described by the following Hamiltonian
\begin{eqnarray}
  H_{t} &=& \frac{1}{2m}\sum_{i=1}^2 \left( \vec{p}_i+\frac{e}{c}
  \vec{A}(\vec{r}_i) \right)^2 + U(\vec{r}_1,\vec{r}_2), \\
  U(\vec{r}_1,\vec{r}_2) &=&
  \frac{1}{2}m\omega_{0}^2 \left( r_1^2 + r_2^2 \right) +\frac{e^2}{\epsilon }
  \frac{1}{\sqrt{|\vec{r}_{1}-\vec{r}_{2}|^{2}+d^{2}}}\label{potential},
\end{eqnarray}
with $m$ the effective electron mass,
$\epsilon $ the dielectric constant of the
medium the electrons are moving in, and $\omega _{0}$ the radial confinement
frequency. A magnetic field is applied perpendicular to the dots,
$\vec{B}=B\vec{e}_z$, which in the symmetric gauge is described by the vector
potential $\vec{A}(\vec{r})=[\vec{B}\times\vec{r}]/2$, and the above Hamiltonian can
be rewritten as follows
\begin{eqnarray}
  H_{t} &=& \frac{1}{2m}\left( p_1^2+p_2^2 \right)
  + \tilde{U}(\vec{r}_1,\vec{r}_2) \nonumber \\
  &&-\frac{1}{2}\omega_c\{[\vec{p}_1\times\vec{r}_1]+[\vec{p}_2\times\vec{r}_2]\}_z,
\end{eqnarray}
where the symbol $\omega_c=eB/mc$ stands for the cyclotron frequency and the
potential
$\tilde{U}$ is given by expression (\ref{potential}) with the scaled confinement
frequency $\tilde{\omega}_0=\gamma\omega_0$ where
$\gamma = \sqrt{1+{\omega_c^2}/{4\omega_0^2}}$.
\par
A crucial property of parabolic confinement potentials is that
the center-of-mass and the relative motion are separable.
The center-of-mass
coordinate and the relative coordinate are $\vec{R}=(\vec{r}_{1}+\vec{r}%
_{2})/2$ and $\vec{r}=\vec{r}_{1}-\vec{r}_{2}$, respectively, with the conjugate
momenta $\vec{P}=\vec{p}_1+\vec{p}_2$ and
$\vec{p}=(\vec{p}_1-\vec{p}_2)/2$. The
center-of-mass and relative motions are described by the following
Hamiltonians  
\begin{mathletters}
\begin{eqnarray}\label{cm}
  W(\vec{P},\vec{R}) &=&\frac{1}{4m}P^{2}+m\omega _{0}^{2}R^{2}
  -\frac{1}{2}\omega_c[\vec{P}\times\vec{R}]_z, \label{cmham}\\
  H(\vec{p},\vec{r}) &=& \frac{1}{m}p^{2}+\tilde{U}(r)
  - \frac{1}{2}\omega_c[\vec{p}\times\vec{r}]_z \label{rmham},
\end{eqnarray}
\end{mathletters}
where
\begin{equation}\label{pot}
  U(r) = \frac{1}{4}m\omega _{0}^{2}r^{2}+\frac{e^{2}}
  {\epsilon }\frac{1}{\sqrt{r^{2}+d^{2}}},
\end{equation}
and $\tilde{U}$ is obtained from $U$ by replacing
$\omega_0$ by $\tilde{\omega}_0$.

In the following we do not have to consider 
the center-of-mass motion because it is a harmonic oscillator with
ground-state energy and excitation frequencies which do not depend on the
interatomic distance $d$. We concentrate our attention on the relative motion.

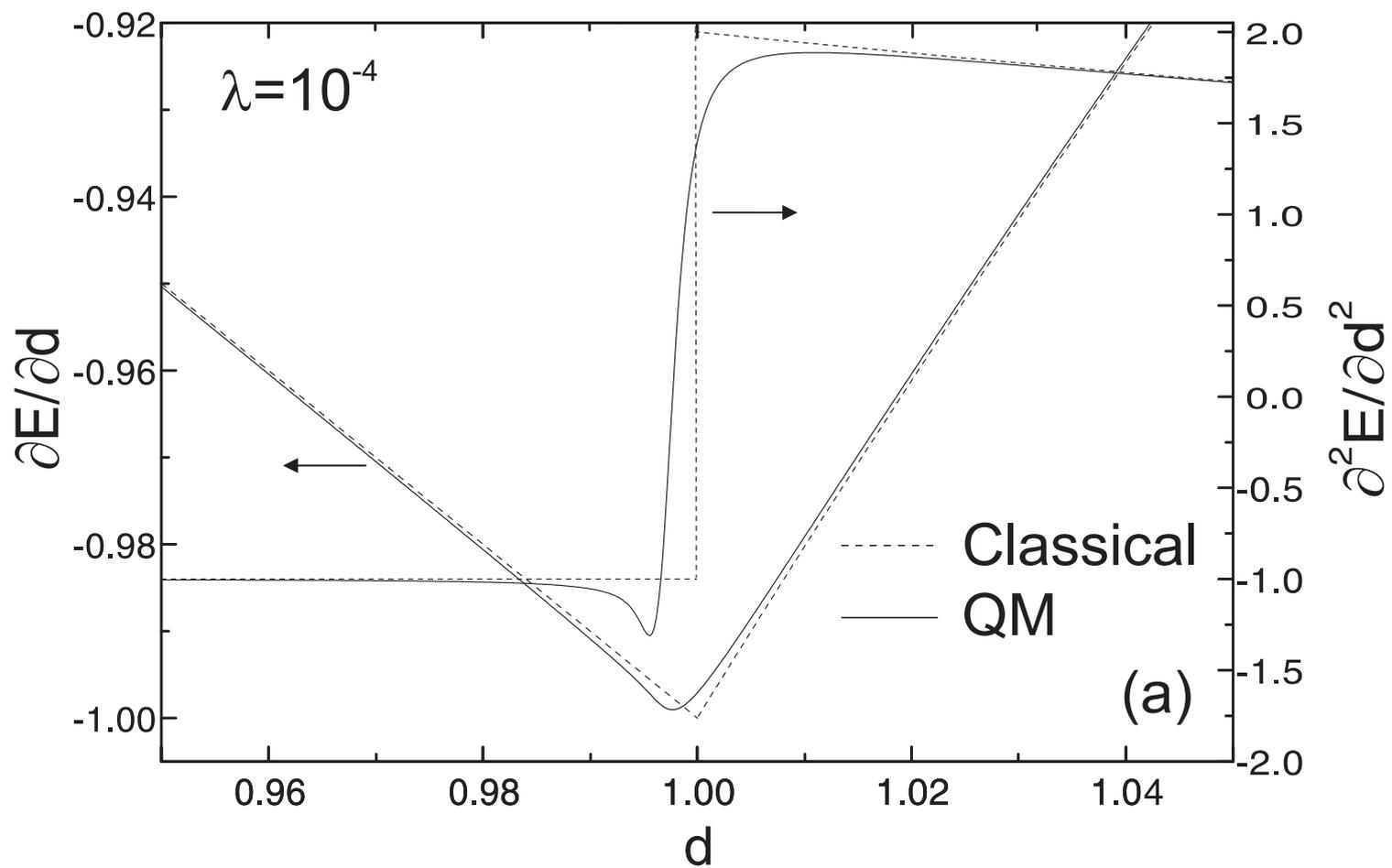
\begin{figure}
\caption{The first and second 
derivative (full curves) of the ground-state 
energy for $\omega_c=0$, i.e.
$n=0,m=0$, for the quantum artificial molecule as a function of $d$ for (a)
$\lambda=10^{-4}$ and (b) $\lambda=0.2$. The classical result ($\lambda=0$) is
shown by the dashed curves.}
\label{fig2}
\end{figure}

%%%%%%%%%%%%%%%%%%%%%%%%%%%%%%%%%%%%%%%%%
%	The Classical solution		%
%%%%%%%%%%%%%%%%%%%%%%%%%%%%%%%%%%%%%%%%%
%\section{The classical solution}
First we consider the classical problem. The Newton equation of motion
for this system is
\begin{equation}\label{equmo}
  \ddot{\vec{r}} = [\fcyc\times\dot{\vec{r}}] - \frac{2}{m}\nabla U(\vec{r}),
\end{equation}
with $\fcyc=\omega_c \vec{e}_z$.
The classical ground-state configuration is obtained by minimizing the potential
$U(\vec{r})$. A second order transistion occurs when the Coulomb interaction
energy and the confinement energy are of the same order. To emphasize this we
write the Hamiltonian in dimensionless form where we express the coordinates
and energy in the following units $r^{\prime }=(e^{2}/\epsilon
)^{1/3}\alpha ^{-1/3}$ and $E^{\prime }=(e^{2}/\epsilon )^{2/3}\alpha
^{1/3}$, respectively, with $\alpha=m\omega_0^2/2$. The potential reduces to
\begin{equation}\label{pot0}
  U(r) = \frac{1}{2}r^{2} + \frac{1}{\sqrt{r^{2}+d^{2}}},
\end{equation}
which after minimalisation 
leads to the equilibrium radius for the ground-state configuration
$r_0(d) = \sqrt{1-d^2}$ for $d<1$, and $r_0(d)=0$ for $d>1$.
The corresponding energy, and its first and
second derivatives with respect to the distance $d$ are $E=(3-d^{2})/2$, $%
\partial E/\partial d=-d$ and $\partial ^{2}E/\partial d^{2}=-1$ for $d<1$,
and $E=1/d$, $\partial E/\partial d=-1/d^{2}$ and $\partial ^{2}E/\partial
d^{2}=2/d^{3}$ for $d>1$. The first and second derivatives are shown in
Fig.~\ref{fig2} by the dashed curves. It is clear
that there is a second order transition at $d=1$. This second order
transition is of the same type as the single second order transitions we
found previously for the artificial molecules with twice six, seven and
eight electrons\cite{psp97}.
\par
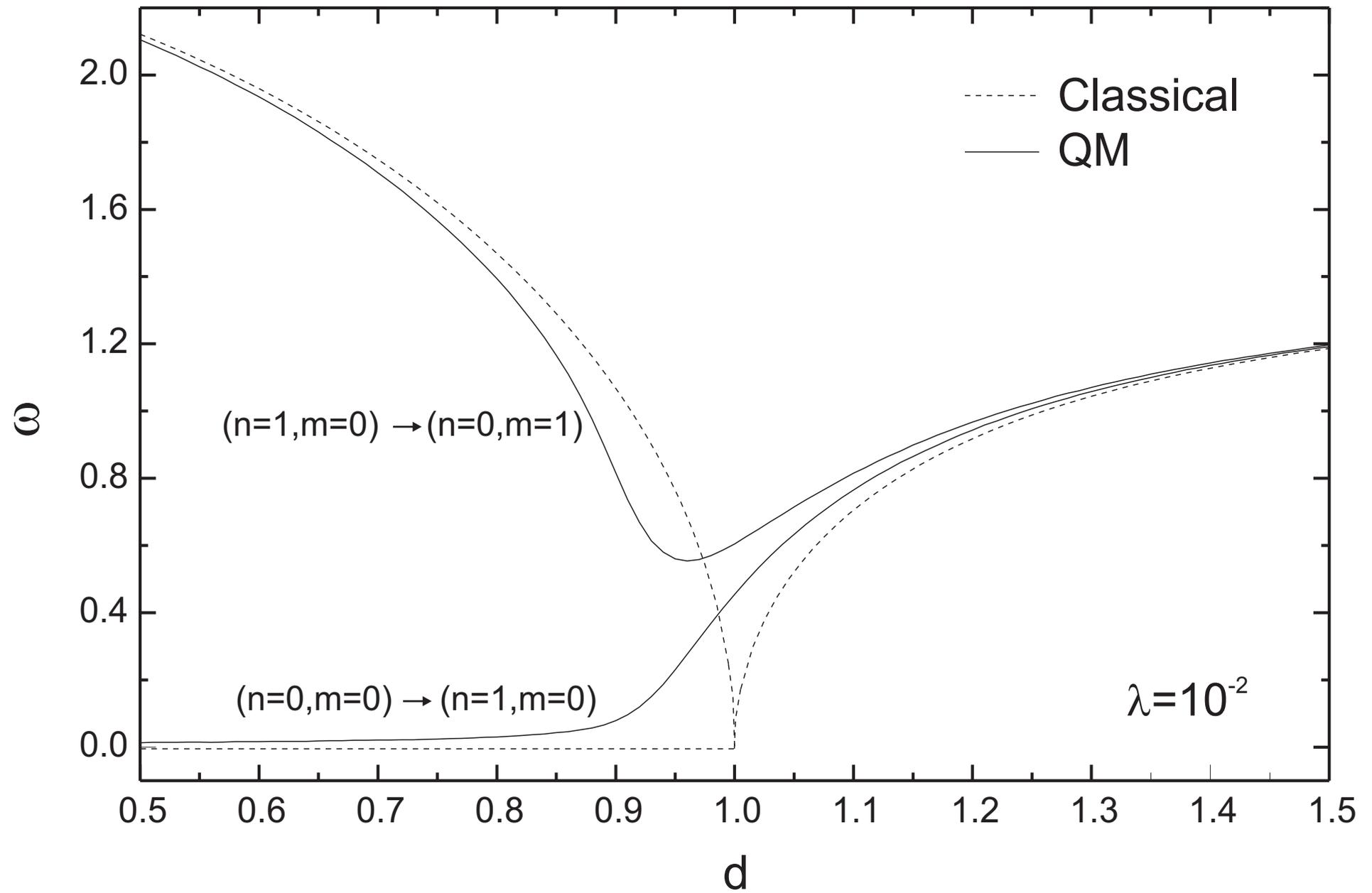
\begin{figure}
\caption{The quantum mechanical frequencies (full curves) for $\lambda=10^{-2}$ 
and the classical frequencies (dashed
curves) as a function of $d$ and for $\omega_c=0$.
%The energy levels, whose
%difference determine the quantum mechanical frequencies, are also given.
}
\label{fig3}
\end{figure}

Next we calculate the frequencies of the eigenmodes. We express
the frequencies in the unit $\omega^{\prime}=\omega_0/\sqrt{2}$ and 
the equation of motion becomes
\begin{equation}\label{equmo0}
  \ddot{\vec{r}} = [\fcyc\times\dot{\vec{r}}] - 2\nabla U(\vec{r}),
\end{equation}
where the dimensionless potential (\ref{pot0}) is used. To find the spectrum
we take the Fourier transform of this Newton equation  
which leads to the following excitation frequencies
\begin{equation}
\begin{array}{ll}
  \omega_r=0,\; \omega_b=\sqrt{6(1-d^2)+\omega_c^2}, & d<1; \\
  \omega_{\pm}=\sqrt{2(1-1/d^3)+(\omega_c/2)^2}\pm\omega_c/2, & d>1.
\end{array}
\end{equation}
The frequency
$\omega_r=0$ corresponds to the rotation of the molecule as a whole, and
$\omega_b$ is the frequency of the breathing mode. The frequencies $\omega_+$
and $\omega_-$ correspond with the out-of-phase motions of the electrons around
the center of the dots.
These classical frequencies are shown in Fig.~\ref{fig3} for $\omega_c=0$ 
by the dashed curves.
Note that one of the eigenmodes softens at $d=1$ which induces the second 
order transition. 

%%%%%%%%%%%%%%%%%%%%%%%%%%%%%%%%%
%	The quantum solution	%
%%%%%%%%%%%%%%%%%%%%%%%%%%%%%%%%%
%\section{The quantum solution}
Now we turn our attention to the quantum mechanical problem and investigate
how quantum effects influence the classical transition. In the quantum case the
ground-state energy can be obtained as a solution of the Schr\"{o}dinger
equation.
%
%\begin{equation}
%  \{H-E\}\Psi=0.
%\end{equation}
%
Due to the cylindrical symmetry of the problem it reduces to the following
equation for the radial wave function ($\Psi(\vec{r})=\exp(-im\varphi)R_m(r)$)
\begin{eqnarray}
  \left\{-\lambda^2 \left(\frac{1}{r}\frac{d}{dr}r\frac{d}{dr}\right)
  +\frac{\lambda^2 m^2}{2}
  -\lambda \frac{\omega_c}{2}m \right.& & \nonumber \\
  \left.+ \tilde{U}(r)-E \right\}R_m(r) &=& 0\label{rad}
\end{eqnarray}
where
\begin{equation}\label{potent}
  \tilde{U}(r) = \frac{\gamma^2}{2}r^2 + \frac{1}{\sqrt{r^2+d^2}},
\end{equation}
and $\lambda = \hbar\omega_0/(\alpha r'^2 \sqrt{2})$, which is the ratio of the
quantum energy with the energy at the transition point.
We solved this
differential equation numerically, using a non uniform space grid as was done
in Ref.~\cite{ps96}.

%%%%%%%%%%%%%%%%%%%%%%%%%
%	Discussion	%
%%%%%%%%%%%%%%%%%%%%%%%%%
%\section{Discussion}
To see how quantum effects influence  the classical transition at $d=1$ we solve
Eq.~(\ref{rad}) for $\omega_c=0$ and we take $\lambda$ slightly different
from 0. In Fig.~\ref{fig2}(a) the first and second
derivative of the ground-state energy (i.e. $n=0,\; m=0$) with respect to $d$ is
given for $\lambda=10^{-4}$, together with the
classical result. Fig.~\ref{fig2}(b) shows the same results
for $\lambda=0.2$. One notices that quantum fluctuations smear out
the classical transition
which happens already for very small values of $\lambda$.
In order to reveal the influence of the quantization on the eigenfrequencies 
obtained classically we define them in the quantum case as the smallest
differences between the excited state energies of Eq.~(\ref{rad}) devided
by $\lambda$. This is shown in Fig.~\ref{fig3} for $\lambda=10^{-2}$. 
In agreement with the disappearance of the
classical transition there is no softening of a frequencymode in the quantum
case.
\par
The effectcs of these quantum fluctuations can be understood from a simple
qualitative consideration.
As we are interested in the behavior of the system close to the transition 
point we
use the following expansion of the potential~(\ref{potent})
\begin{equation}
  \tilde{U}(r) = \frac{1}2 \left( \gamma^2-\frac{1}{d^3}
  \right) r^2 +\frac{3r^4}{8d^5} \sim
  \frac{3}{2}\Delta d\; r^2 + \frac{3}{8}r^4,
\end{equation}
where we neglect the constant shift and introduced the parameter
$\Delta d = (d - \tilde{d}_0)\gamma^{8/3}$, which
characterizes the deviation from the exact transistion point
$\tilde{d}_0 = \gamma^{-2/3} \approx 1-{\omega_c^2}/{12\omega_0}$.
Thus in the quantum case the magnetic field shifts the transition
point to lower $d$ values. 

After the above expansion we arrive at the following radial equation which describes the quantum behavior
of
the system for small $\lambda$ values and close to the transition point
\begin{eqnarray}
  \left\{-\lambda^2 \left(\frac{1}{r}\frac{d}{dr}r\frac{d}{dr}\right)
  +\frac{\lambda^2 m^2}{r^2} \right.&&\nonumber \\
  \left.+ \frac{3}{2}\Delta d\; r^2 + \frac{3}{8}r^4 \right. -&& \left. E 
  \right\}R_m(r) = 0.\label{rad0}
\end{eqnarray}
Note that there are two parameters $d$ and $\lambda$ which control the quantum
problem while in the classical case there is only one parameter $d$ which controls the
closeness to the transition point. The role of the other parameter $\lambda$
can be understood by  considering the following limiting cases, where we take
$\omega_c=0$ for simplicity.
In the case when the parameter $\Delta d$ is not too small
one may neglect the quartic term in Eq.~(\ref{rad0}) and we obtain the 
Schr\"{o}dinger equation for a harmonic
oscillator. This
leads to the following excitation frequency estimation
$\omega \sim E/ \lambda \sim \sqrt{\Delta d}$. Consequently,
not too close to the transition point the excitation frequency
demonstrates the classical soft mode behavior.
In the opposite limiting case close to the transition point (small $\Delta d$
values)
the quadratic term can be neglected. Scaling the variable
$r\to\lambda^{1/3}r$
it can be estimated that the excitation frequencies are 
of order $\omega \sim E/\lambda \sim \lambda^{1/3}$. 
Therefore, in the
quantum case the soft mode frequency does not tend
to zero, as is also apparent in Fig.~\ref{fig3}. Comparing the frequency 
estimations in both $\Delta d$ regions we find that in the region
$|\Delta d|<\lambda^{2/3}$
quantization influences the classical transition point behavior. 
For $\lambda=10^{-4}$,
as in Fig.~\ref{fig2}(a), this region is about $\Delta d\approx 10^{-2}$, which
agrees with the numerical results of
Fig~\ref{fig2}(a).
\par
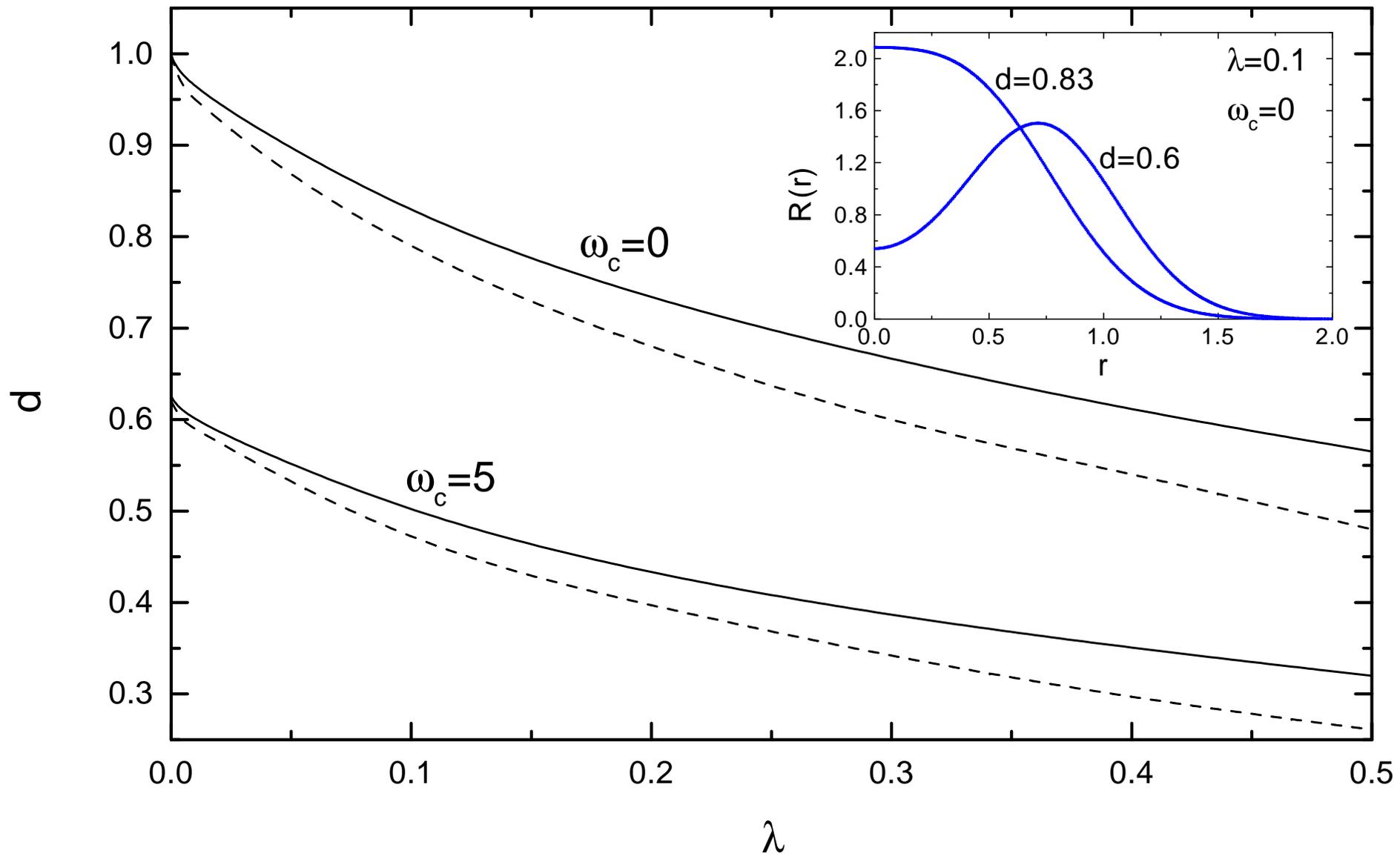
\begin{figure}
\caption{The dashed curve is the $d$ value for
wich $\partial E/\partial d$ is minimal, which is a measure for the shift
of the transition. The full curve is the minimum $d$ value at which the radial
wavefunction (inset of figure) attains a maximum at $r=0$.
The results are shown for $\omega_c=0$ and $\omega_c=5$.} 
\label{fig4}
\end{figure}

From Fig.~\ref{fig2} we also notice that in the quantum case 
the transition is shifted to
smaller $d$ values. As a measure for this shift we take the
interatomic distance
value at which $\partial E/\partial d$ attains its minimum value for the
ground-state energy ($n=0,\;m=0$). This 
interatomic distance (thick curve in Fig.~\ref{fig4})
coincides almost with the $d$ value (dashed curve in Fig.~\ref{fig4}) 
at which the radial wavefunction (inset of Fig.~\ref{fig4}) becomes
maximal in the center of the dot. 
The shift of the transition to smaller $d$ values can also be induced by 
applying an external magnetic field (see Fig.~\ref{fig4}). 
The plotted results for $\omega_c=5$ are also for the state ($n=0,\;
m=0$), which is not the ground-state anymore, except for large $d$ values.
Thus in the quantum case the electrons sit in the center of both dots for
smaller $d$ values than in the classical case, which causes the shift to lower
$d$ values.

%%%%%%%%%%%%%%%%%%%%%%%%%%%%%%%%%%%%%%%%%%%%%%%%%%%%%%%%%
%	SPIN SINGLET SPIN TRIPLET TRANSITION		%
%%%%%%%%%%%%%%%%%%%%%%%%%%%%%%%%%%%%%%%%%%%%%%%%%%%%%%%%%
%\section{Spin-singlet $\leftrightarrow$ spin-triplet transitions}
%\par
%Now we consider the quantum problem in the presence of a magnetic field
%perpendicular to the dots. 
It is known that in quantum dots~\cite{wmc92} and in the $D^-$
problem~\cite{rsp98} oscillations between
spin-singlet and spin-triplet ground-states occur as a function of the magnetic
field strength. Here we want to investigate how these 
oscillations change
when the distance between the dots increases and how they are altered or
modified by the presence of the classical transition. 
%The origin of these spin-singlet $\leftrightarrow$ spin-triplet oscillations is the
%electron-electron interaction. When the magnetic field is increased the orbits
%of the electrons converge to the center of the dot, which increases the Coulomb
%energy. Therefore, for a particular value of the magnetic field it can be energitically
%more favourable to increase the angular momentum ($m \rightarrow m+1$): this
%increases the rotational energy, but lowers the Coulomb energy. At that point
%the sign of the radial wavefunction changes, and thus also the sign of the spin
%part. This is the spin-singlet $\leftrightarrow$ spin-triplet transition. 
It is 
%then also 
obvious that when the interatomic distance $d$ increases, a higher magnetic
field is needed to induce these transitions
and that for sufficiently large $d$ no transitions occur.
\par
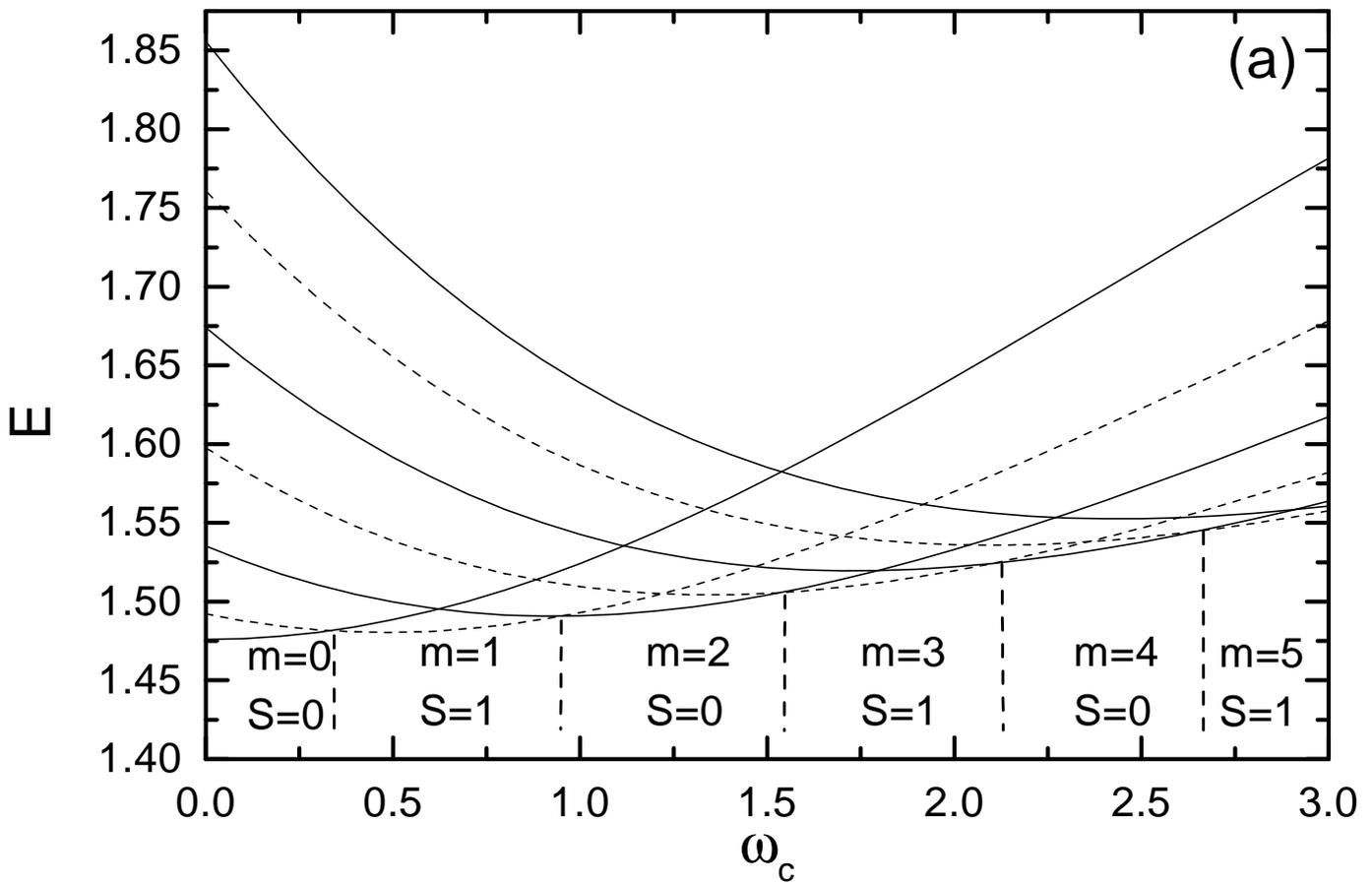
\begin{figure}
\caption{(a) The lowest energy levels for the two-electron double quantum dot
problem ($d=0.5, \lambda=0.1$). The singlet phase $m=0$ ($S=0$) at low
magnetic fields is followed by
$m=1$ ($S=1$), $m=2$ ($S=0$), $\ldots$ phases as the magnetic field strength is
increased.
(b) Phase diagram for singlet and triplet ground-states for $\lambda=0.1$.
%The singlet phase $m=0$ ($S=0$) in zero and low magnetic fields is followed by
%$m=1$ ($S=1$), $m=2$ ($S=0$), $\ldots$ phases as the magnetic field strength is
%increased. 
Only the first eight transitions are shown.
When $d>1$ no transitions occur.}
\label{fig5}
\end{figure}

In Fig.~\ref{fig5}(a) these spin-singlet $\leftrightarrow$ spin-triplet oscillations are shown
for $d=0.5$ and $\lambda=0.1$.
In Fig.~\ref{fig5}(b) we plot the phase diagram 
for the system with $\lambda=0.1$. When $d$ increases, the
transitions are shifted to higher magnetic fields. The classical transition is
seen as follows: for $d>1$ we found 
no spin-singlet $\leftrightarrow$ spin-triplet transitions anymore, 
while for $d<1$ oscillations do occur
and this for all values of $\lambda$. As in the
classical system, the interlayer 
correlations are sufficiently strong to impose
the one atom properties for $d<1$, and for $d>1$ the
system behaves like two decoupled quantum dots.
The spin-singlet $\leftrightarrow$ spin-triplet oscillations in quantum dots 
are related to the Wigner crystallization state of the quantum
dot~\cite{maksym96,anisim98} what in our case
corresponds to the transition at $d=1$ and which explains why for $d>1$ those
oscillations are absent.

Acknowledgments. B.P. is a research assistant and F.M.P. a research director
with the Flemisch Science Foundation (FWO-Vlaanderen). Part of this work is
supported by IUAP-IV and FWO-Vlaanderen.

%
%\begin{figure}
%\caption{The results for the classical artificial molecule as a function of the
%interatomic distance $d$. (a) The distance of
%the electrons from the center of  the confinement potential. 
%(b) The energy, its first and second derivative with respect to $d$. 
%(c) The frequencies of the classical eigenmodes for $\omega_c=0$ and
%$\omega_c=1$.}
%\label{fig1}
%\end{figure}
%

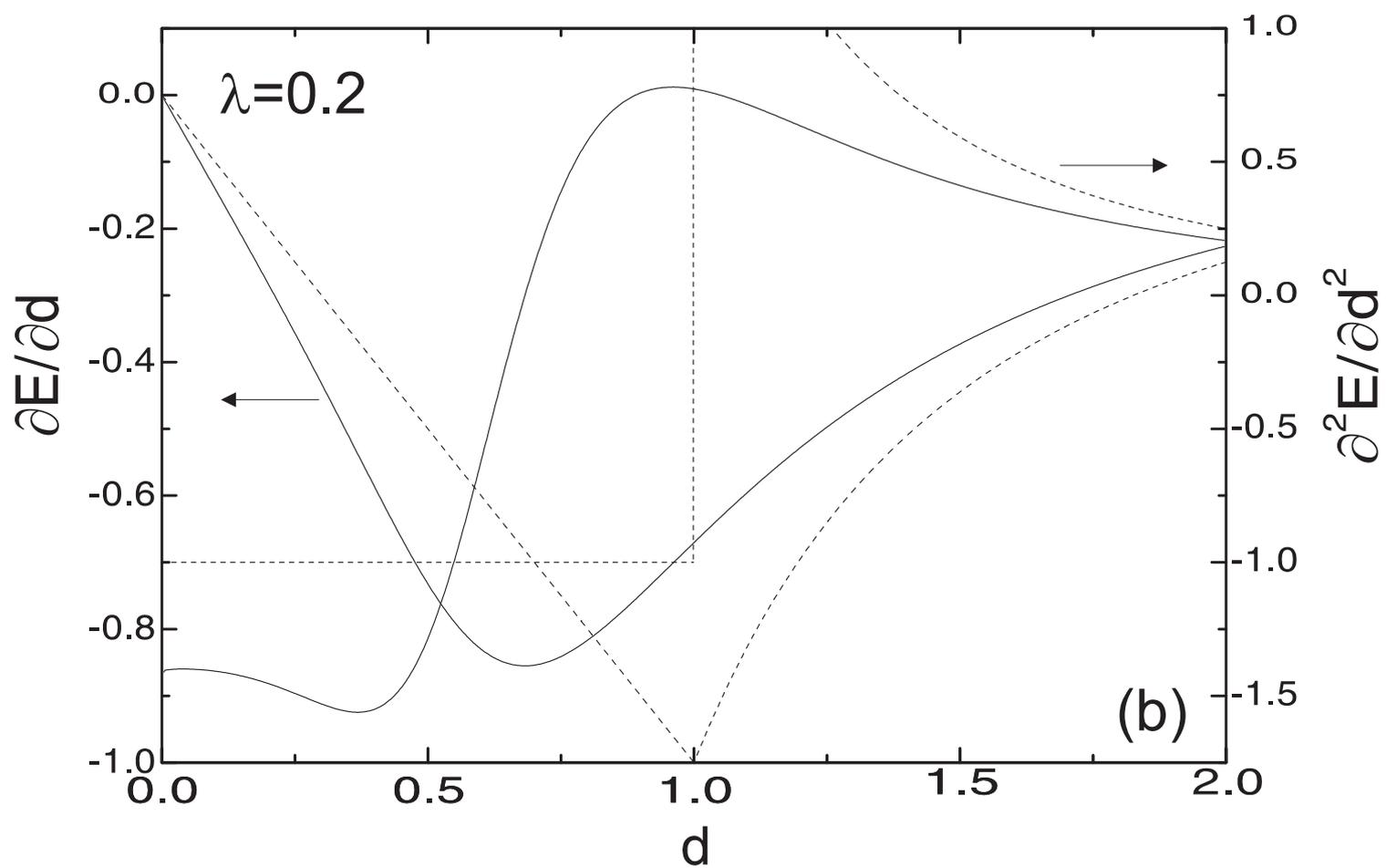
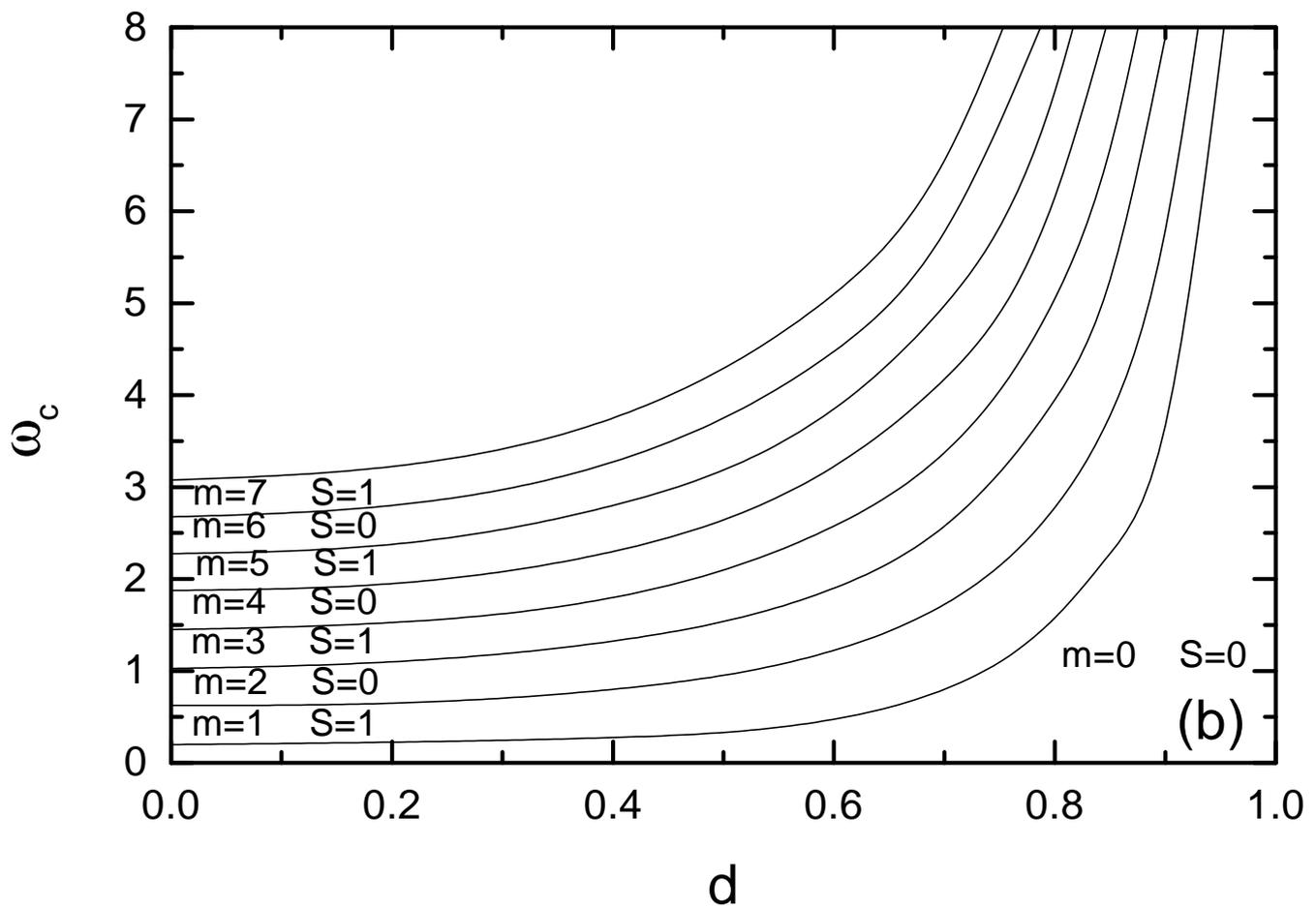
\end{document}